\documentclass[11pt]{article}
\usepackage{fullpage}

\usepackage{graphicx}
\usepackage{url}
\usepackage{comment}
\usepackage{booktabs}
\usepackage{listings}

\usepackage[noend]{algorithmic}
\usepackage{algorithm}

\def\th#1{\multicolumn{1}{c}{{\sf #1}}}  % table header

\title{Roomy: A System for Space Limited Computations}

\author{Daniel Kunkle \\
Northeastern University, Boston, USA \\
\url{kunkle@ccs.neu.edu}}

\date{}

\begin{document}

\maketitle

\begin{abstract}
There are numerous examples of problems in symbolic algebra in which the
required storage grows far beyond the limitations even of the distributed RAM
of a cluster.  Often this limitation determines how large a problem one can
solve in practice.  Roomy provides a minimally invasive system to modify the
code for such a computation, in order to use the local disks of a cluster or a
SAN as a transparent extension of RAM.

Roomy is implemented as a C/C++ library. It provides some simple data
structures (arrays, unordered lists, and hash tables). Some typical
programming constructs that one might employ in Roomy are: map, reduce,
duplicate elimination, chain reduction, pair reduction, and breadth-first
search.  All aspects of parallelism and remote I/O are hidden within the Roomy
library.
\end{abstract}

\section{Introduction}

This paper provides a brief introduction to Roomy~\cite{roomy}, a new programming
model and open source library for parallel disk-based computation.  The
primary purpose of Roomy is to solve space limited problems without
significantly increasing hardware costs or radically altering existing
algorithms and data structures.

Roomy uses disks as the main working memory of a computation, instead of RAM.
These disks can be disks attached to a single shared-memory system,  a storage
area network (SAN), or the locally attached disks of a compute cluster.
Particularly in the case of using the local disks of a cluster, disks are often
underutilized and can provide several order of magnitude more working memory
than RAM for essentially no extra cost.

There are two fundamental challenges in using disk-based storage as main memory:

\begin{itemize}

\item[] {\bf Bandwidth}: roughly, the bandwidth of a single disk is 50 times less
than that of a single RAM subsystem (100 MB/s versus 5 GB/s). The solution is
to use many disks in parallel, achieving an aggregate bandwidth comparable to
RAM.

\item[] {\bf Latency}: even worse than bandwidth, the latency of disk is
many orders of magnitude worse than RAM. The solution is to avoid latency
penalties by using streaming data access, instead of costly random access.

\end{itemize}

Roomy hides from the programmer both the complexity inherent in parallelism
and the techniques needed to convert random access patterns into streaming
access patterns.  In doing so, the programming model presented to the user
closely resembles that of traditional RAM-based serial computation.

The rest of this paper briefly describes the data structures provided by Roomy,
and some example programming constructs that can be implemented using these
data structures. Complete documentation, and instructions for obtaining the Roomy
open source library, can be found on the Web at \url{roomy.sourceforge.net}.

\section{Roomy Data Structures}

Roomy data structures are transparently distributed across many disks, and the
operations on these data structures are transparently parallelized across the
many compute nodes of a cluster. Currently, there are three Roomy data
structures:

\begin{itemize}

\item[] {\tt RoomyArray}: a fixed size, indexed array of elements (elements can
be as small as one bit).

\item[] {\tt RoomyHashTable}: a dynamically sized structure mapping {\em keys} to
{\em values}.

\item[] {\tt RoomyList}: a dynamically sized, unordered list of elements.

\end{itemize}

There are two types of Roomy operations: delayed and immediate.  If
an operation requires random access, it is delayed. Otherwise, it is performed
immediately. To initiate the processing of delayed operations for a given Roomy
data structure, the programmer makes an explicit call to {\em synchronize} that
data structure. By delaying random access operations they can be collected and
performed more efficiently in batch.

Table~\ref{tab:ops} describes some of the basic Roomy operations. Some
operations are specific to one type of Roomy data structure, while others apply
to all three. The operations are also identified as either immediate (I) or
delayed (D).

For performance reasons, it is often best to use a {\tt RoomyArray} or {\tt
RoomyHashTable} instead of a {\tt RoomyList}, where possible.  Computations
using {\tt RoomyList}s are often dominated by the time to sort the list and any
delayed operations. {\tt RoomyArray}s and {\tt RoomyHashTable}s avoid sorting
by organizing data into {\em buckets}, based on indices or keys.

\begin{table*}
\centering

\caption{Some basic Roomy operations. If an operation is specific to one type
of data structure, it is listed under {\tt RoomyArray}, {\tt RoomyHashTable},
or {\tt RoomyList}. Otherwise, it is listed as ``common to all''.  Also, the
type of each operation is given as either immediate (I) or delayed (D).}

\vspace{1ex}
\begin{tabular}{llcp{0.55\textwidth}}
\toprule
\th{Data Structure} & \th{Name} & \th{Type} & \th{Description} \\
\midrule
{\tt RoomyArray}
 & {\tt access} & D & apply a user-defined function to an element \\
 & {\tt update} & D & update an element using a user-defined function \\
\\

{\tt RoomyHashTable}
 & {\tt insert} & D & insert a given (key, value) pair in the table\\
 & {\tt remove} & D & given a key, remove the corresponding (key, value) pair
                      from the table \\
 & {\tt access} & D & given a key, apply a user-defined function to the
                      corresponding value \\
 & {\tt update} & D & given a key, update a the corresponding value using
                      a user-defined function \\
\\

{\tt RoomyList}
 & {\tt add}         & D & add a single element to the list \\
 & {\tt remove}      & D & remove all occurrences of a single element from the
                           list \\
 & {\tt addAll}      & I & adds all elements from one list to another \\  
 & {\tt removeAll}   & I & removes all elements in one list from another \\
 & {\tt removeDupes} & I & removes duplicate elements from a list \\
\\

Common to all
 & {\tt sync}   & I & process all outstanding delayed operations for the
                      data structure \\
 & {\tt size}   & I & returns the number of elements in the data structure \\
 & {\tt map}    & I & applies a user-defined function to each element \\
 & {\tt reduce} & I & applies a user-defined function to each element and returns
                      a value (e.g. the ten largest elements of the list) \\
 & {\tt predicateCount} & I & returns the number of elements that satisfy a given
                              property (Note: this does not require a separate
                              scan, the count is kept current as the data is
                              modified) \\
\bottomrule
\end{tabular}
\label{tab:ops}
\end{table*}

\section{Programming Constructs}

Because Roomy provides data structures and operations similar to traditional 
programming models, many common programming constructs can be implemented
in Roomy without significant modification. The one major difference is in the
use of delayed random operations. To ensure efficient computation, it is important
to maximize the number of delayed random operations issued before they are
executed (by calling {\tt sync} on the data structure).

Below are Roomy implementations of six programming constructs: map, reduce, set
operations, chain reduction, pair reduction, and breadth-first search. Both map
and reduce are primitive operations in Roomy. The others are built using Roomy
primitives.

\newpage

First, note that the code given here uses a simplified syntax. For example, the
{\tt doUpdate} method from the {\em chain reduction} programming construct
below would be implemented in Roomy as:

\begin{lstlisting}[language=C,frame=single]
void doUpdate(uint64 localIndex, void* localVal,
              void* remoteVal) {
  *(int*)localVal =
      *(int*)localVal + *(int*)remoteVal;
}
\end{lstlisting}

The simplified version given here eliminates the type casting, and appears as:

\begin{lstlisting}[language=C,frame=single]
int doUpdate(int localIndex, int localVal,
             int remoteVal) {
  return localVal + remoteVal;
}
\end{lstlisting}

A future C++ version of Roomy is planned that would use templates to make the
simplified version legal code.

See the online Roomy documentation and API~\cite{roomy} for the exact syntax
and function definitions.

\paragraph{Map}

The {\tt map} operator applies a user-defined function to every element of a
Roomy data structure.  As an example, the following converts a {\tt RoomyArray}
into a {\tt RoomyHashTable}, with array indices as keys and the associated
elements as values.

\begin{lstlisting}[language=C,frame=single]
RoomyArray ra;      // elements of type T
RoomyHashTable rht; // pairs of type (int, T)

// Function to map over RoomyArray ra.
void makePair(int i, T element) {
  RoomyHashTable_insert(rht, i, element);
}

// Perform map, then complete delayed inserts
RoomyArray_map(ra, makePair);
RoomyHashTable_sync(rht);
\end{lstlisting}

\paragraph{Reduce}

The {\tt reduce} operator produces a result based on a combination of all
elements in a data structure. It requires two user-defined functions. The first
combines a partially computed result and an element of the list. The second
combines two partially computed results.  The order of reductions is not
guaranteed. Hence, these functions must be associative and commutative, or else
the result is undefined.

As an example, the following computes the sum of squares of the elements in a
{\tt RoomyList}.

\begin{lstlisting}[language=C,frame=single]
RoomyList rl;  // elements of type int

// Function to add square of an element to sum.
int mergeElt(int sum, int element) {
  return sum + element * element;
}

// Function to compute sum of two partial answers.
int mergeResults(int sum1, int sum2) {
  return sum1 + sum2;
}

int sum =
    RoomyList_reduce(rl, mergeElt, mergeResults);
\end{lstlisting}

The type of the result does not necessarily have to be the same as the
type of the elements in the list, as it is in this case. For example, the
result could be the $k$ largest elements of the list.

\paragraph{Set Operations}

Roomy can support certain set operations through the use of a {\tt RoomyList}.
Some of these operations (particularly intersection) are sub-optimal when
built using the current set of primitives. Future work is planned to add
a native {\tt RoomySet} data structure.

A {\tt RoomyList} can be converted to a set by removing duplicates.

\begin{lstlisting}[language=C,frame=single]
RoomyList A;  // can contain duplicate elements
RoomyList_removeDupes(A);  // now a set
\end{lstlisting}

Performing set union, $A = A \cup B$, is also simple.

\begin{lstlisting}[language=C,frame=single]
RoomyList A, B;
RoomyList_addAll(A, B);
RoomyList_removeDupes(A);
\end{lstlisting}

Set difference, $A = A - B$, is performed by using just the {\tt removeAll}
operation, assuming $A$ and $B$ are already sets.

\begin{lstlisting}[language=C,frame=single]
RoomyList A, B;
RoomyList_removeAll(A, B);
\end{lstlisting}

Finally, set intersection is implemented as a union, followed set differences:
$C = (A + B) - (A - B) - (B - A)$. Set intersection may become a Roomy
primitive in the future.

\begin{lstlisting}[language=C,frame=single]
// input sets
RoomyList A, B;
// initially empty sets
RoomyList AandB, AminusB, BminusA, C;

// create three temporary sets
RoomyList_addAll(AandB, A);
RoomyList_addAll(AandB, B);
RoomyList_removeDupes(AandB);
RoomyList_addAll(AminusB, A);
RoomyList_removeAll(AminusB, B);
RoomyList_addAll(BminusA, B);
RoomyList_removeAll(BminusA, A);

// compute intersection
RoomyList_addAll(C, AandB);
RoomyList_removeAll(C, AminusB);
RoomyList_removeAll(C, BminusA);
\end{lstlisting}

\paragraph{Chain Reduction}

Chain reduction combines each element in a sequence with the element after it.
In this example, we compute the following function for an array of integers
{\tt a} of length {\tt N}

\begin{lstlisting}[language=C,frame=single]
for i = 1 to N-1
  a[i] = a[i] + a[i-1]
\end{lstlisting}

\noindent where all array elements on the right-hand side are accessed before
updating any array elements on the left-hand side.

In the following code, {\tt val\_i} represents {\tt a[i]} and \\
{\tt val\_iMinus1} represents {\tt a[i-1]}.

\begin{lstlisting}[language=C,frame=single]
RoomyArray ra;  // array of ints, length N

// Function to complete updates
int doUpdate(int i, int val_i, int val_iMinus1) {
  return val_i + val_iMinus1;
}

// Function to be mapped over ra, issues updates
void callUpdate(int iMinus1, int val_iMinus1) {
  int i = iMinus1 + 1;
  if i < N
    RoomyArray_update(
        ra, i, val_iMinus1, doUpdate);
}

RoomyArray_map(ra, callUpdate); // issue updates
RoomyArray_sync(ra);         // complete updates
\end{lstlisting}

The computation is deterministic. The new array values will be based only on
the old array values because Roomy guarantees that none of the delayed update
operations will be executed until {\tt sync} is called.  The code above is
implemented internally through a traditional scatter-gather operation.

\paragraph{Parallel Prefix}

The chain reduction programming construct can also be used as the basis for a
parallel prefix computation. At a high level, the parallel prefix computation
is defined as

\begin{lstlisting}[language=C,frame=single]
for (k = 1; k < N; k = k * 2)
  if i-k >= 0
    a[i] = a[i] + a[i-k];
\end{lstlisting}

\paragraph{Pair Reduction}

Pair reduction applies a function to each pair of elements in a collection.
For an array {\tt a} of length {\tt N}, pair reduction is defined as

\begin{lstlisting}[language=C,frame=single]
for i = 0 to N-1
  for j = 0 to N-1
    f(a[i], a[j]);
\end{lstlisting}

The following example inserts each pair of elements from a {\tt RoomyArray}
into a {\tt RoomyList}. The variable {\tt outerVal} represents {\tt a[i]} and
the variable {\tt innerVal} represents {\tt a[j]}.

\begin{lstlisting}[language=C,frame=single]
RoomyArray ra;  // array of int, length N
RoomyList rl;   // list containing Pair(int, int)

// Access function, adds a pair to  the list
void doAccess(int innerIndex, int innerVal,
              int outerVal) {
  RoomyList_add(
      rl, new Pair(innerVal, outerVal));
}

// Map function, sends access to all other elts
void callAccess(int outerIndex, int outerVal) {
  for innerIndex = 0 to N-1
    RoomyArray_access(
        ra, innerIndex, outerVal, doAccess);
}

RoomyArray_map(ra, callAccess);
RoomyArray_sync(ra);  // perform delayed accesses
RoomyList_sync(rl);   // perform delayed adds
\end{lstlisting}

One can think of the {\tt RoomyArray\_map} method as the outer loop,
the {\tt callAccess} method as the inner loop, and the {\tt doAccess} method 
as the function being applied to each pair of elements.

\paragraph{Breadth-first Search}

Breadth-first search enumerates all of the elements of a graph, exploring
elements closer to the starting point first. In this case, the graph is
implicit, defined by a starting element and a generating function that returns
the neighbors of a given element.

\begin{lstlisting}[language=C,frame=single]
// Lists for all elts, current, and next level
RoomyList* all = RoomyList_make("allLev", eltSize);
RoomyList* cur = RoomyList_make("lev0", eltSize);
RoomyList* next = RoomyList_make("lev1", eltSize);

// Function to produce next level from current
void genNext(T elt) {
  /* User-defined code to compute neighbors ... */
  for nbr in neighbors
    RoomyList_add(next, nbr);
}

// Add start element
RoomyList_add(all, startElt);
RoomyList_add(cur, startElt);

// Generate levels until no new states are found
while(RoomyList_size(cur)) {
  // generate next level from current
  RoomyList_map(cur, genNext);
  RoomyList_sync(next);

  // detect duplicates within next level
  RoomyList_removeDupes(next);

  // detect duplicates from previous levels
  RoomyList_removeAll(next, all);

  // record new elements
  RoomyList_addAll(all, next);
  
  // rotate levels
  RoomyList_destroy(cur);
  cur = next;
  next = RoomyList_make(levName, eltSize);
}
\end{lstlisting}

One of the initial tests of Roomy was to use breadth-first search to solve the
{\em pancake sorting problem}. Pancake sorting operates using a sequence of
prefix reversals (reversing the order of the first $k$ elements of the
sequence). The sequence can be thought of as a stack of pancakes of varying
sizes, with the prefix reversal corresponding to flipping the top $k$ pancakes.
The goal of the computation is to determine the number of reversals required to
sort any sequence of length~$n$.

Using Roomy, the entire application took less than one day of programming and less
than 200~lines of code. Breadth-first search was implemented using a {\tt
RoomyArray}, similar to the {\tt RoomyList}-based version presented above.
% It was able to solve the 13-pancake problem in 5~hours using the locally
% attached disks of a 30~node cluster.

Three different solutions to the pancake sorting problem, each using one of the
three Roomy data structures, is available in the Roomy online
documentation~\cite{roomy}.

\section{Acknowledgments}

Many thanks to Gene Cooperman and Vlad Slavici, whose input has greatly
improved both Roomy and this paper.

\bibliographystyle{plain}
\bibliography{ref-kunkle}

\end{document}